\documentclass{nature}

\usepackage{siunitx}
\usepackage{graphicx} 
\usepackage{amsmath}

\newcommand*\diff{\mathop{}\!\mathrm{d}}

\title{Self-wrapping of an Ouzo Drop Induced by Evaporation on a Superamphiphobic Surface}

\author{Huanshu Tan$^{1}$, Christian Diddens$^{2}$, Michel Versluis$^{1}$, Hans-Jürgen Butt$^{3}$, Detlef Lohse$^{1,4}$ \& Xuehua Zhang$^{5,1}$}

\begin{document}

\maketitle

\begin{affiliations}
 \item Physics of Fluids group, Department of Science and Technology, Mesa+ Institute, and J. M. Burgers Centre for Fluid Dynamics, University of Twente, P.O. Box 217, 7500 AE Enschede, The Netherlands,
  \item Department of Mechanical Engineering, Eindhoven University of Technology, P.O. Box 513, 5600 MB Eindhoven, The Netherlands,
  \item Max Planck Institute for Polymer Research, 55128 Mainz, Germany,
  \item Max Planck Institute for Dynamics and Self-Organization, 37077 Gottingen, Germany,
  \item Soft Matter \& Interfaces Group, School of Engineering, RMIT University, Melbourne, VIC 3001, Australia.
\end{affiliations}

\begin{abstract}
Evaporation of multi-component drops is crucial to various technologies and has numerous potential applications because of its ubiquity in nature.
Superamphiphobic surfaces, which are both superhydrophobic and superoleophobic, can give a low wettability not only for water drops but also for oil drops.
In this paper, we experimentally, numerically and theoretically investigate the evaporation process of millimetric sessile ouzo drops (a transparent mixture of water, ethanol, and trans-anethole) with low wettability on a superamphiphobic surface.
The evaporation-triggered Ouzo effect, i.e. the spontaneous emulsification of oil microdroplets below a specific ethanol concentration, preferentially occurs at the apex of the drop due to the evaporation flux distribution and volatility difference between water and ethanol. This observation is also reproduced by numerical simulations.
The volume decrease of the ouzo drop is characterized by two distinct slopes. The initial steep slope is dominantly caused by the evaporation of ethanol, followed by the slower evaporation of water. 
At later stages, thanks to Marangoni forces the oil wraps around the drop and an oil shell forms.
We propose an approximate diffusion model for the drying characteristics, which predicts the evaporation of the drops in agreement with experiment and numerical simulation results.
This work provides an advanced understanding of the evaporation process of ouzo (multi-component) drops.\end{abstract}

\section{Introduction}
Drop evaporation is an omnipresent phenomenon in daily life.
During this process, the liquid at the drop surface changes its phase and escapes as vapor into the ambient air.
Dating back to Maxwell, the evaporation process of a drop in an ambient gas has been explored and considered mainly as a diffusion-controlled process \cite{maxwell1890}.
The study of the evaporation process of sessile drops is important because of its crucial role in numerous technologies and applications, such as inkjet printing, coatings, patternings, deposition of materials, or DNA mapping \cite{picknett1977,deegan1997,hu2002,popov2005,dash2013,semenov2013,paul2001,park2006,Kim2004,Schena467,bensimon1995,Sirringhaus2123,Erbil201267,cazabat2010evaporation,tagkey2015i}.
In the last two decades, numerous studies have focused on understanding the evaporation process of sessile drops on solid substrates experimentally, numerically and theoretically \cite{Erbil201267,cazabat2010evaporation}.
Surface properties \cite{birdi1993,Kulinich2009,Anantharaju2009,erik2015}, thermal effects \cite{hu2005,pan2013}, dispersed particles in the liquid \cite{deegan1997,Yunker2011}, surfactants at the liquid-gas interface \cite{still2012,Sempels2013,marin2016}, and the liquid composition \cite{liu2008,sefiane2008,christy2011} were all found to have a contribution to the drop evaporation characteristics.

The evaporation of multi-component drops draws special interest because of its ubiquity in practice.
The physicochemical properties of the drop solution dramatically enrich the system and give rise to an unexpected outcome:
the different volatilities of the components lead to distinct evaporation stages with different evaporation rates and various types of wetting behavior \cite{liu2008,sefiane2008}.
In a hydrodynamic context, flow transitions inside an evaporating binary mixture drop have been revealed, which are a result of the intense and complicated coupling of flow and the spatio-temporal concentration field \cite{christy2011,bennacer2014}.
By controlling these mechanisms, binary drop evaporation can offer a new physicochemical way for surface coatings \cite{Kim2016a}.

Recently, we used an ouzo drop as a model for a \textit{ternary} liquid mixture and investigated its evaporation process on a hydrophobic surface \cite{tanouzo2016}.
The Greek drink Ouzo (or the French Pastis or the Turkish Raki) is a miscible solution and primarily consists of water, ethanol and anise oil.
When the water concentration is increased by adding water or by reducing ethanol, the solution becomes opaque due to the ``ouzo effect'', i.e. the spontaneous nucleation of oil microdroplets \cite{Katz2003liquid,ganachaud2005nanoparticles,botet2012ouzo}.
We discovered how the preferential evaporation of ethanol triggers the ``ouzo effect'' in an evaporating ouzo drop.
Four life phases can be distinguished during its drying. 
As a remarkable phenomenon, we found that the evaporation-triggered nucleation starts at the rim of the ouzo drops.
It can be attributed to the drop geometry, namely the singularity at the rim.
On the hydrophobic substrate, the ouzo drops maintain a contact angle $\theta$ smaller than \SI{90}{\degree} (flat droplets) because of the low surface energy of ethanol in the ouzo solution.
The singularity of the evaporation rate at the contact line of flat drops \cite{popov2005,stauber2015evaporation} and the higher volatility of ethanol induce a maximum in the local water concentration and thereby trigger the onset of the ``ouzo effect'' at this position.
At a later stage, the microdroplets coalesce and form an oil ring at the rim of the drop, with a water drop sitting on it. 
Once also the water has evaporated, only an oil drop remains.

Based on the diffusion model by Popov \cite{popov2005}, changing the geometric configuration is a simple and direct way to change the evaporation profile and hence to induce a different concentration distribution along the liquid-air interface. 
In particular, a maximum evaporation rate should be found at the top of the drop when the drop has a large contact angle ($\theta>$ \SI{90}{\degree}).
Thus, ouzo drops on substrates with low wettability should have the highest water concentration at the top of the drop rather than at the rim, and therefore the onset of nucleation should take place right there.
However, a comprehensive numerical model by Pan \textit{et al.}, who took into account the evaporative cooling effect and the buoyancy-driven convective flow in the drop and vapor domains, shows that the maximum evaporation rate of drops with low wettability is still located at the contact line due to temperature effects \cite{pan2013}.
Both of these models are only applicable to \textit{single-component} drops, but hitherto it little is known about the evaporation process of multi-component sessile drops with low wettability ($\theta>$ \SI{90}{\degree}).
Here we explore where the evaporation-triggered nucleation process starts for evaporating ouzo drops with $\theta>$ \SI{90}{\degree}, and find out the evaporation dynamics.

First, we present an investigation of the evaporation characteristics of millimetric ouzo drops on a flat surface with a large contact angle. 
We performed evaporation experiments on a superamphiphobic surface, which is both superhydrophobic and superoleophobic, to achieve low wettability for the ouzo drops ($\theta>$ \SI{150}{\degree} to start).
We found that the ``ouzo effect'' induced by evaporation indeed preferentially takes place at the top of the drop, and two distinct stages with different evaporation rates exist. 
Moreover, a new remarkable phenomenon appears: part of the nucleated oil microdroplets form an oil shell \textit{wrapping up} the ouzo drop, instead of forming a persistent oil ring at the contact line.
Then a numerical model based on a finite element method presents additional insight into the process.
Finally, we propose an approximate diffusion model for the evaporation characteristics of ouzo (multi-component) drops, and furthermore highlight and discuss the influences of Marangoni flow and the evaporative cooling effect.
In summary, we provide a quantitive understanding of the evaporation process of ouzo drops experimentally, numerically and theoretically.

\section{Materials and Methods}
\subsection{Ouzo Droplet Solution and Superamphiphobic Substrate }
The ouzo drop solution was prepared with an initial composition of \SI{32.02}{\percent} (vol/vol) Milli-Q water [produced by a Reference A+ system (Merck Millipore) at 18.2 M$\Omega\cdot$cm (at \SI{25}{\degreeCelsius})], \SI{66.5}{\percent} (vol/vol) ethanol (SIGMA-ALDRICH; $\geq$ \SI{99.8}{\percent}) and \SI{1.48}{\percent} (vol/vol) trans-anethole (SIGMA-ALDRICH; \SI{99}{\percent}).
In this work, we used pure trans-anethole, which is the main component of anise oil, as the oil phase in ouzo drop to rule out any influence from the other components in anise oil.
Experiments were carried out on a soot-templated superamphiphobic glass substrate \cite{Deng2012,paven2014}.
These soot-templates surfaces are formed by collecting a fractal-like network of
self-assembled nearly spherical carbon particles (diameter about \SI{40}{\nano\meter}). The
network is roughly \SI{30}{\micro\meter} thick and homogeneous on length scales above \SI{5}{\micro\meter}.
The soot particles are loosely connected by van der Waals forces. The
network was stabilized by depositing roughly \SI{30}{\nano\meter} of SiO2 using chemical
vapor deposition (CVD) of tetraethyl orthosilicate (TEOS) for \SI{24}{\hour}. The
final porosity was \SI{90}{\percent} \cite{paven2014}.
The soot-templated superamphiphobic surface is optically
transparent, so that the bottom side of the ouzo drop can be imaged while
it evaporates.
The static contact angles of Milli-Q water and trans-anethole oil on the substrate are \SI{160}{\degree} $\pm$ \SI{1}{\degree} and \SI{157}{\degree} $\pm$ \SI{0.5}{\degree}, respectively.
Thus the ouzo drop with more than 60\% ethanol can still initially hold $\sim150^{\circ}$ static contact angle.

 \subsection{Experimental Setup}
 
 \begin{figure*}[ht]
 \centering
 \includegraphics[width=.95\linewidth]{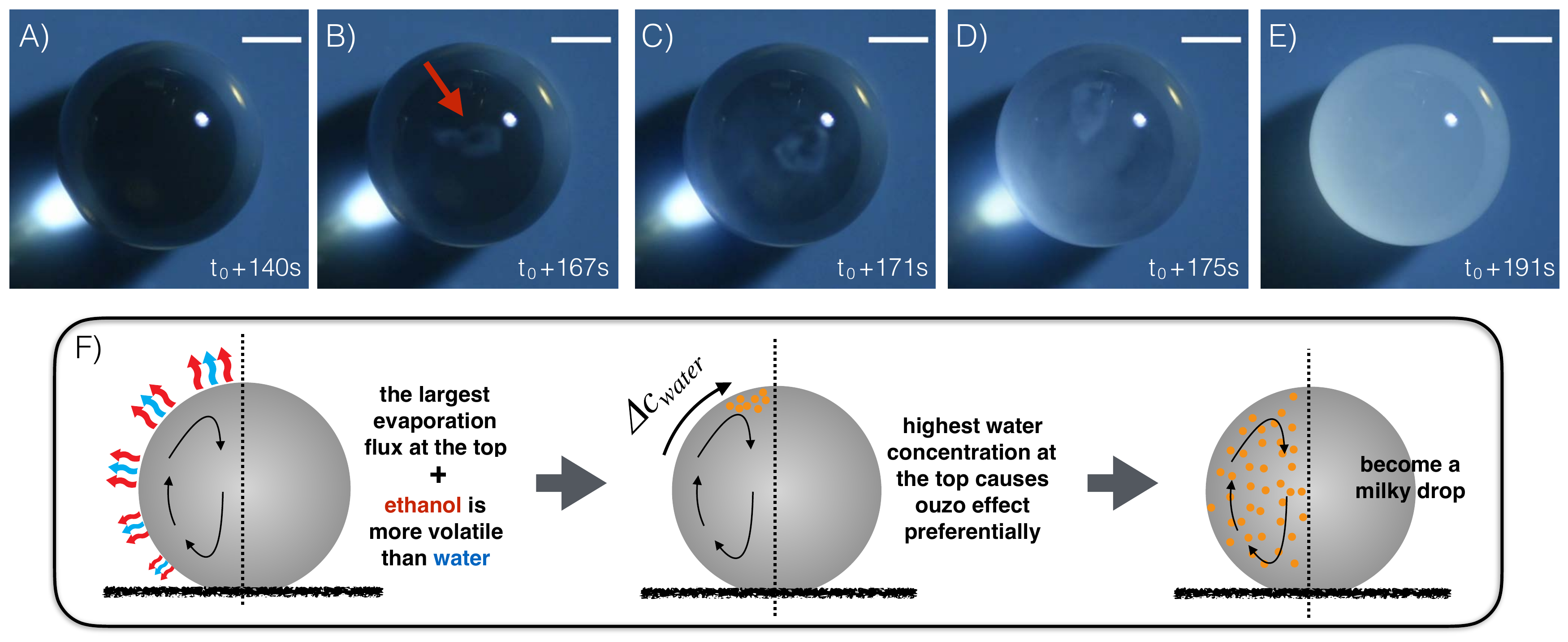}
 \caption{A-E) Experimental top view snapshots during the evaporation-triggered nucleation at the top of an evaporating ouzo drop on a flat superamphiphobic surface. The scale bar is \SI{0.5}{\milli\metre}.
Originally, the ouzo drop is transparent in A. The light point is the focal point.
At some point later in B, a region of cloudy white emulsions (see red arrow) appears at the drop top due to evaporation-triggered nucleation. The light scattering by the nucleated microdroplets leads to the white color of the emulsions.
In C and D, the white emulsions are more and more visible and start to spread around the entire drop.
Finally, the entire drop is opaque in E.
F) Explanation diagram of experimental snapshots A-E. 
The distribution of the evaporation flux and the different volatilities of water and ethanol codetermine that ``ouzo effect '' preferentially happens at the drop apex. Marangoni flow spreads the nucleated oil microdroplet through the entire drop.}
 \label{fgr:snapshots}
\end{figure*}
 
The evaporation experiments were performed in an empty room without any person in the room during data recording. In this sense, the entire lab can be considered as closed cell.
An ouzo drop was deposited on a superamphiphobic surface through a teflonized needle (HAMILTON; 8646-02) by a motorized syringe pump (HARVARD; PHD 2000).
Experiment recording started when the needle departed from the drop (defined as $t_0$).
In practice there was a time delay ($\sim$ \SI{22}{\second}) between starting to pump liquid out of the needle and taking the first snapshot of the drop ($t_0$), leading to premature evaporation.
The time delay was caused by the difficulty of depositing a sessile ouzo drop on a superamphiphobic surface.
Therefore, the initial ratios of ethanol and water of the recording data should be corrected.
The correction for the ethanol composition can be estimated by extrapolating the ethanol composition of the prepared solution (\SI{66.5}{\percent} vol/vol) with respect to the delayed time and the initial volume loss rate (ref to Fig.~\ref{fgr:model}A).
With this method, a \SI{-7.4}{\percent} correction is applied to the initial ethanol composition (\SI{+7.4}{\percent} to water) for the data used in sections~\ref{sec:num} and \ref{sec:model}.
The entire evaporation process of the ouzo drop was recorded by a CCD camera [XIMEA; MD061MU-SY, 3 frames per second (fps) at $1,372 \times 1,100$ pixel resolution] equipped with a high-magnification zoom lens system (THORLABS; MVL12X3Z) for side-view recordings and a CMOS [NIKON; D750, 24 frames per seconds (fps) at $1,920 \times 1,080$ pixels resolution] attached to an identical lens system for top-view recordings.
We used a self-built collimated LED source system to illuminate the side-view recording.
A powerful Hella LED light source was used for the top-view illumination to show the top of the drop (Figs.~\ref{fgr:snapshots}A-E).
The relative humidity and temperature in the laboratory were monitored at a sampling rate of one per second with a universal handheld test instrument (OMEGA; HH-USD-RP1, accuracy relative humidity is $\pm$ \SI{2}{\percent} over 10 to \SI{90}{\percent} @\SI{25}{\degreeCelsius} and a temperature accuracy  of $\pm$ \SI{0.3}{\kelvin} @\SI{25}{\degreeCelsius}).
The location of the probe was around 10 cm away from the droplet.
A similar sketch of the setup is described in detail in reference \cite{tanouzo2016}.
Stereo-imaging was performed by a confocal microscope (Nikon Confocal Microscopes A1 system) with a $20\times$ air objective (CFI Plan Apochromat VC $20\times/0.75$ DIC, NA = 0.75, WD = 1.0 mm).
Perylene (SIGMA-ALDRICH; sublimed grade, $\geq$ \SI{99.5}{\percent}) was used to label trans-anethole.
The presented 3D images have a resolution of \SI{1.2}{\micro\metre} in the horizontal plane and a vertical resolution of \SI{700}{\nano\metre}.
We applied the pendant drop method to measure the interfacial surface tension between different two phases (air, water or trans-anethole) with a video-based optical contact angle measuring system (DataPhysics OCA15 Pro).

\subsection{Image Analysis and Data Calculation}
Images were analyzed by a custom made MATLAB program.
The initial contact line position of the drop was used as input for each data group and automatically adjusted during the subsequent frames.
The program fits an ellipse to the contour of the drop in side view. The contact angle $\theta$ was calculated based on position of the intersection of the base line and the fitted ellipse for each frame.
The volume $V$ of the drop was obtained by integrating the areas of horizontal disk layers with the assumption that each horizontal layer fulfills rotational symmetry.

\section{Experimental Results}
\subsection{Evaporation-triggered Nucleation at the Top of an Evaporating Ouzo Droplet}
\label{sec:nucleEXP}

Figures~\ref{fgr:snapshots}A-E display an evaporating ouzo drop with low wettability on a superamphiphobic substrate under ambient conditions.
Initially, cf. Figure~\ref{fgr:snapshots}A, the ouzo drop is transparent and concentrates illuminating light on the substrate.
Around $t=t_0+$1\SI{67}{\second}, the ``ouzo effect'' sets in \textit{at the top of the drop} and a region with a cloudy white emulsion appears (Fig.~\ref{fgr:snapshots}B).
The nucleated microdroplets scatter the light, giving them ``milky'' appearance.
In Figures~\ref{fgr:snapshots}CD, the emulsion is more and more evident and spreading around the drop.
Finally, the entire drop is opaque, and the bright spot on the substrate disappears completely (Fig.~\ref{fgr:snapshots}E).
Experimental movies (SV1 and SV2) and numerical movie (SV3) are available as supplementary material.

The physical origin of the phenomenon has two aspects. One is the maximum local evaporation rate at the top of the drop. 
On a superamphiphobic substrate, the ouzo drop maintains its high contact angle ($\sim$ \SI{150}{\degree}) during the evaporation process.
Under the assumption of a small temperature gradient along the liquid-air interface, this geometric configuration gives the highest local evaporation flux at the top of the drop \cite{popov2005,stauber2015evaporation,cazabat2010evaporation} (Fig.~\ref{fgr:snapshots}F).
The thermal gradient along the liquid-air interface is reduced by the strong solutal and thermal Marangoni convection, as discussed in detail in section~\ref{sec:marangoni}.
The average temperature at the interface is lower than ambient temperature due to evaporative cooling (section~\ref{sec:coolingeffect}).

The second aspect is that the component ethanol in the ouzo drop has a higher volatility than water, while trans-anethole is non-volatile.
As the drop evaporates, the highest water concentration initially appears at the drop top with the highest local evaporation flux.
Hence, the evaporation-triggered ouzo effect \cite{tanouzo2016} starts at the apex of the ouzo drop.
The nonuniform concentration field induces surface tension gradient along the interface towards the top, which leads to intense solutal Marangoni convection.
The flow drives nucleated oil microdroplets around the drop.
After some time, the oil microdroplets nucleate in the whole drop.
In the end, numerous oil microdroplets fill up the ouzo drop and scattering the light, creating the milky appearance of the drop.

\subsection{Evaporation Phases}
\begin{figure*}[ht]
 \centering
 \includegraphics[width=.99\linewidth]{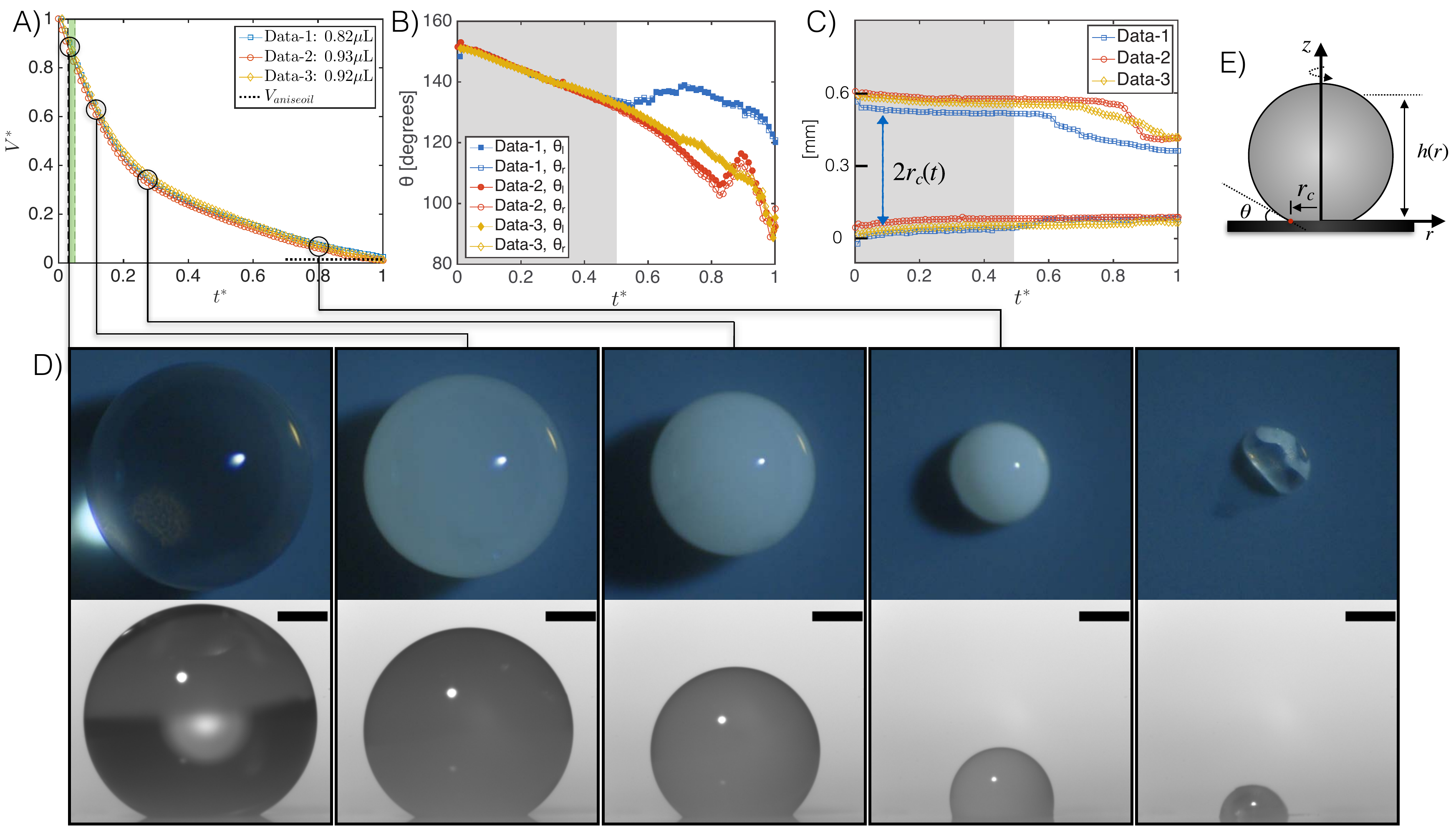}
 \caption{A) Temporal evolution of the ouzo drop volume during the evaporation on the surperamphiphobic surface for different initial drop sizes. The drop volumes are nondimensionalized with respect to the initial volume, while the time is nondimensionalized with respect to the total time of water and ethanol evaporation. Two distinct volume evolution slopes are evident. 
The time interval marked green indicates approximately the transition of the ouzo drops from transparent to opaque.
B) Temporal evolution of the contact angle. For each data group, left and right contact angles are in good agreement, i.e. no bias, at an early stage (grey region), while different evolutions appear afterwards.
C) Temporal evolution of the size of the contact area, characterized by $2r_c(t)$. For the ouzo drops in our case the CR-model is roughly applicable at the early stage (grey region).
D) Experimental snapshots (up: top view; bottom: side view)\textit{$^{\S}$}  at different moments related to panel A. The scale bar is \SI{0.25}{\milli\metre}. The last group snapshot shows the residual drop after evaporation recording. Only oil is left.
E) The sketch of an ouzo drop with annotations.}
 \label{fgr:evp}
\end{figure*}

\footnotetext{\S~With respect to the top view snapshots in Fig.2D, the side-view recording was recorded from the right side by the camera placed upside down. This was due to the equipment installation and the setup arrangement in our experiments.}
\begin{figure*}[ht]
 \centering
 \includegraphics[width=.75\linewidth]{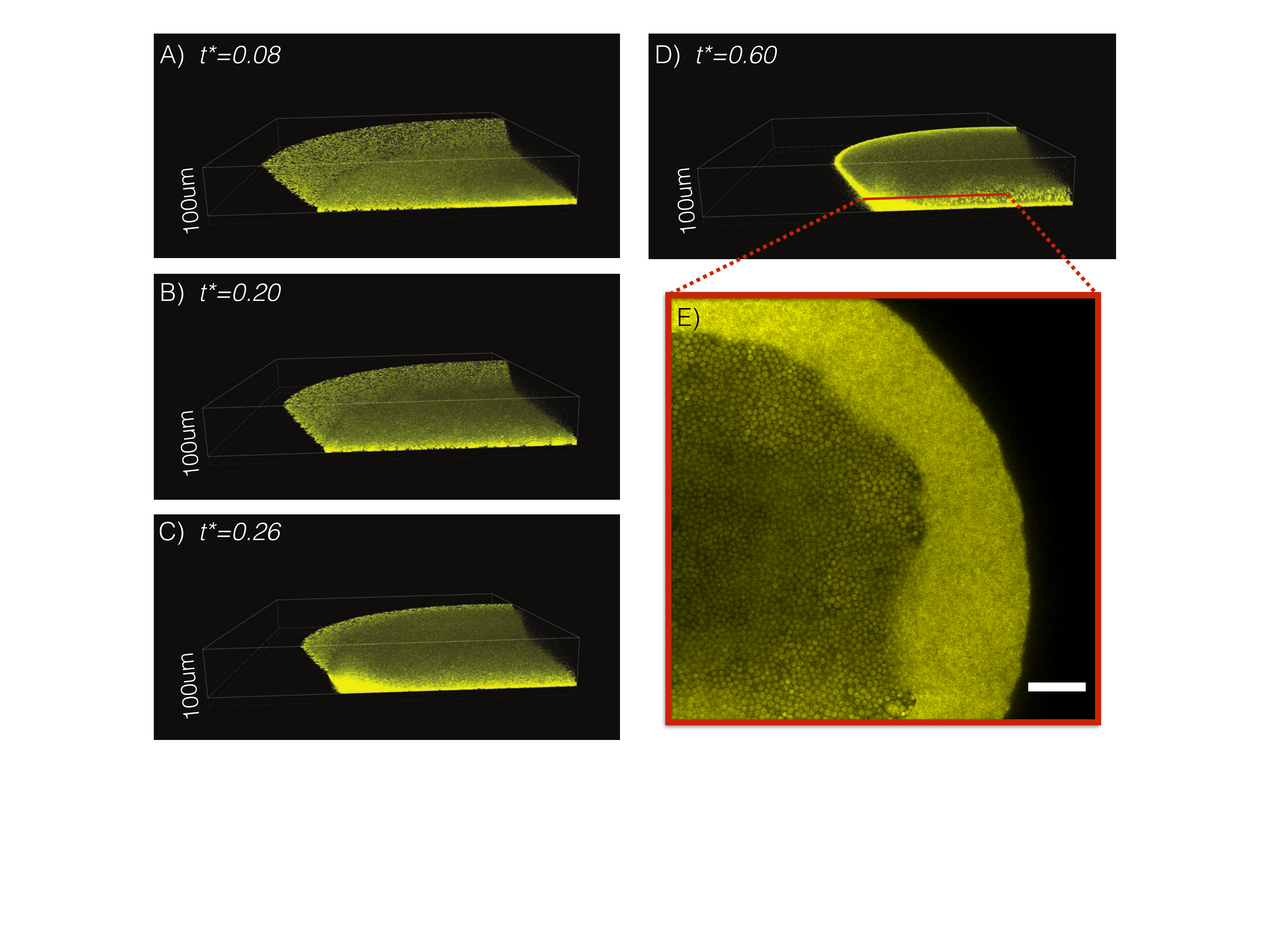}
 \caption{Wrapping of nucleated oil in the evaporating ouzo drop.
 A-D) Three-dimensional confocal images of the contact region of the ouzo drop at different moments with 20$\times$ air objective.
 After oil nucleation, nucleated oil microdroplets (yellow) fill up the ouzo drop. The microdroplets on the bottom and near the surface are much more visible than the ones inside because of the direct laser excitation without light refraction. In panels A and B, as the ouzo drop evaporates, the contact angle decreases with apparent movement of the contact line. In C, the accumulated oil microdroplets fuse into a temporary oil ring at the rim. After a while ($t^*=0.6$ and ethanol has almost entirely evaporated), an oil shell wraps up the ouzo drop, as shown in D.
E) A high resolution two-dimensional confocal snapshot inside the drop with a 60$\times$  oil immersion objective.
Numerous nucleated oil microdroplets in water phase surroundings are capsuled by the oil shell.
The scale bar is \SI{20}{\micro\metre}.
The microdroplets have a diameter size around $\sim$ \SI{2}{\micro\metre}.}
 \label{fgr:wrapping}
\end{figure*}

We monitored evaporating ouzo drops with initial volumes $V_{init}$ of 0.82, 0.93 or \SI{0.92}{\micro\liter} to study its evaporation characteristics, including the transient drop volume $V$, the contact angle $\theta$, and the radius of the contact line $r_c$ (as annotated in Figure~\ref{fgr:evp}E).
Here, we nondimensionalized the temporal evolution of the drop volume as follows:
we define a nondimensional volume $V^*$ by dividing the drop transient volume $V$ by the original amount $V_{init}$.
A nondimensional time $t^*$ for the first three phases is determined by dividing time $t$ by the total time of the water/ethanol evaporation time ($\sim$ \SI{660}{\second}, $\sim$ \SI{727}{\second} and $\sim$ \SI{736}{\second} for Data-1, Data-2, and Data-3, respectively).
After the non-dimensionalization, the three temporal evolution curves overlap, as shown in Figure~\ref{fgr:evp}A.

Like on flat hydrophobic surfaces \cite{tanouzo2016} two distinct time regimes could be
distinguished for drops evaporating from superamphiphobic surfaces (Fig.~\ref{fgr:evp}A).
The similar feature also exists for an evaporating water-ethanol drop in gas \cite{liu2008,sefiane2008,christy2011} and a dissolving sessile drop in an ouzo system \cite{song2016}.
The difference of evaporation/diffusion rates of the components in the drop determines this feature.
In the first period ($t^*<0.2$) of the volume evolution curves in Figure~\ref{fgr:evp}A, 
the ouzo drop undergoes a transition from phase I to phase II, i.e. from transparent (first snapshot column in Fig.~\ref{fgr:evp}D) to opaque (second snapshot column in Fig.~\ref{fgr:evp}D).
For the ouzo drops in Data-1, Data-2 and Data-3, the onset of the transition is at $\sim \SI{21}{\second}, \sim \SI{21}{\second}$ and $\sim \SI{23}{\second}$ after being deposited on the surface, respectively. The transitions take $\sim \SI{12}{\second}$, $\sim \SI{12}{\second}$, and $\sim \SI{17}{\second}$ accordingly to finish the transition. The transition interval (green region in Fig.~\ref{fgr:evp}A) is short compared to the whole evaporation process.
The steep decrease of the initial volume results from the high volatililty of ethanol.
During this period, the contact angle smoothly drops from $\sim$ \SI{150}{\degree} to $\sim$ \SI{140}{\degree} (Fig.~\ref{fgr:evp}B), whereas the contact radius $r_{c}$ has a subtle shift (Fig.~\ref{fgr:evp}C), i.e. the drop is roughly in the Constant Radius (CR)-mode \cite{lohse2015rmp}.
We note that not only the left and right contact angles of a single drop coincide, but also the contact angles of different drops.
Furthermore, three-phase intersection points in the side-view contour, which were reported in our previous study on flat evaporating ouzo drops \cite{tanouzo2016}, are now absent.

When most of the ethanol has evaporated, the evaporation of water dominates and determines the less steep slope of phase III.
The ouzo drop is milky during the whole phase III (the third and fourth snapshot columns in Fig.~\ref{fgr:evp}D). 
There is no phase inversion from oil droplets in water to water droplets in oil as was reported for flat evaporating ouzo drops \cite{tanouzo2016}.
The nucleated oil emulsion microdroplets fill up the ouzo drop and remain stable.
After some time, the contact angle increases and the contact area simultaneously shrinks to a smaller base radius (Figs.~\ref{fgr:evp}BC), 
i.e. the CR-mode ceases.
As the water is evaporating, its concentration in the drop continues to decrease until there is not enough water to maintain the stability of oil emulsion microdroplets in the bulk. The microdroplets then start to coalesce (last snapshot column in Fig.~\ref{fgr:evp}D), leaving behind a small trans-anethole drop.
The capillary force at the contact line on the surface can damage the soot coating layer. The residual trans-anethole drop rolls up the damaged layer, resulting in a non-spherical-cap shape in the end.

\subsection{Wrapping of Nucleated Oil}
As discussed before, the drop initially is in CR-mode \cite{lohse2015rmp} in the early stages of the evolution (for example, $t^*<0.5$ for Data-1).
To further investigate this behavior, we used a confocal microscope to image the contact region of an evaporating ouzo drop.
Figures~\ref{fgr:wrapping} AB show the nucleated oil microdroplets (yellow) in the bulk or on the surface. 
The presence of numerous oil microdroplet leads to a high light absorption, which shields the oil microdroplets inside the ouzo drop. 
Thanks to the visibility of oil microdroplets along the drop surface, the behavior of the drop surface in the contact region is detectable.
The confocal images confirm that at early time ($t^*\leq 0.2$) the contact angle of the drop decreases and the contact line is fixed.
At some point later, inward movement of the contact line facilitates the coalescence of the nucleated oil microdroplet on the surface, forming a temporary oil ring as displayed in Figure~\ref{fgr:wrapping}C.

The oil ring does not persist for a long time.
It starts to climb along the liquid-air interface and wraps up the ouzo drop at some point when the drop has a small ethanol concentration and a high surface energy.
Smith \textit{et al.} \cite{smith2013} and Schellenberger \textit{et al.} \cite{frank2015} have reported the same phenomenon before.
The spreading coefficient gives a criterion for the occurrence of the wrapping process \cite{smith2013}.
It is defined as $S_{\text{ta,l}} \equiv \gamma_{\text{l,a}}-\gamma_{\text{ta,a}}-\gamma_{\text{l,ta}}$, where $\gamma$ is the interfacial tension between different phases, l is short for the ouzo drop liquid, a is for air, and ta is for trans-anethole \cite{de2004capillarity} .
If $S_\text{ta,w}$ is positive, the trans-anethole-air interface and trans-anethole-liquid together have a lower energy than the liquid-air interface and thus trans-anethole can wrap up the drop \cite{kontogeorgis2016}.
Table~\ref{tbl:s} lists the interfacial tension between water and air ($\gamma_{\text{w,a}}$), trans-anethole and air ($\gamma_{\text{ta,a}}$), and water and trans-anethole ($\gamma_{\text{w,ta}}$).
If the drop liquid is water, the spreading coefficient $S_{\text{ta,w}}$ is positive.
Consequently, the spreading parameter predicts total wetting and the trans-anethole wraps up the ouzo drop completely.
As shown in Figure~\ref{fgr:wrapping}D, a bright oil shell appears.
An experimental movie (SV4) created by confocal microscope is available as supplementary material.
To have an observation inside the ouzo drop at high resolution, we performed a 2D scan with an $60 \times$ oil immersion objective in the confocal microscope system.
Figure~\ref{fgr:wrapping}E shows a snapshot, where we find numerous nucleated oil microdroplets with $\sim$ \SI{2}{\micro\metre} diameter size by the  continuous oil shell.
Here, only fluorescent dye was added for trans-anethole.

\begin{table*}
\small
\centering
  \caption{\ Interfacial surface tension between different two phases: $\text{a}$ is short for air, $\text{w}$ for water, and $\text{ta}$ for trans-anethole.}
  \label{tbl:s}
  \begin{tabular*}{0.5\textwidth}{@{\extracolsep{\fill}}c c c c}
    \hline
    $\gamma_{\text{w,a}}$ [\SI{}{\milli\newton}/\SI{}{\metre}] & $\gamma_{\text{ta,a}}$ [\SI{}{\milli\newton}/\SI{}{\metre}] & $\gamma_{\text{w,ta}}$ [\SI{}{\milli\newton}/\SI{}{\metre}] & $S_{\text{ta,w}}$ [\SI{}{\milli\newton}/\SI{}{\metre}] \\
    \hline
    72 & 35.5 & 24.2 & \text{$>$}0 \\
    \hline
  \end{tabular*}
\end{table*}

\section{Numerical Modeling of the Evaporation Process with a Finite Element Method}
\label{sec:num}
Additional insight in the entire process can be obtained by numerical modeling. Since the initial contact angle is higher than \SI{90}{\degree}, the lubrication theory model of references \cite{tanouzo2016,Diddens2017a} cannot be used. To overcome this limitation and to provide an accurate prediction of the flow velocity even at high contact angles, a new finite element method (FEM) model has been developed. Here, we only give an outline of this model, and details can be found in reference \cite{Diddens2017b}. In order to allow for acceptable calculation times, the model assumes axisymmetry. Furthermore, it is assumed that the drop is always in a spherical-cap shape and consists of a miscible liquid mixture. When the ouzo effect occurs, the latter assumption is still valid as long as the oil microdroplets are small compared to the entire drop. In the simulations, the ouzo effect is defined to happen when the local composition is in the experimentally determined ouzo effect regime of the ternary phase diagram of reference \cite{tanouzo2016}.

\begin{figure*}[ht]
\centering
\includegraphics[width=\textwidth]{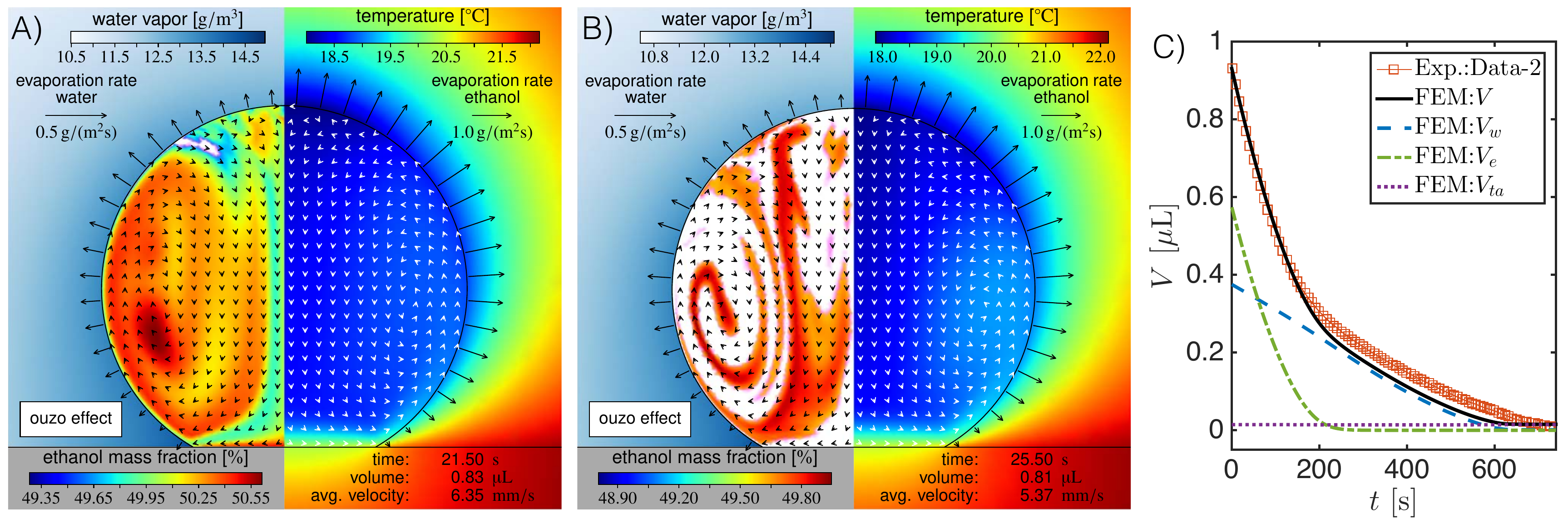}
\caption{Snapshots of the simulation at different times, $t=\SI{21.5}{\second}$ (A) and $t=\SI{25.5}{\second}$ (B), and good agreement between volume evolutions from experiments and numerical simulations (C). Snapshots A-B) The left side shows water vapor concentration $c_\text{w}$ in the gas phase and ethanol mass fraction $y_\text{e}$ in the droplet and the right side presents temperature field. The arrows in the drop indicate the flow direction and the evaporation rates $j_{\text{w}}$ and $j_{\text{e}}$ of water and ethanol are indicated by the interface arrows on the left and right side, respectively. At about $t=\SI{21.5}{\second}$, the ouzo effect occurs at the apex of the drop (indicated by white regions on the left side).  At about $t=\SI{27.2}{\second}$, the phase separation is occurring in the entire drop. 
C) shows good agreement of volume evolutions of experimental data (square symbols) and simulations (black solid line). During the first 200 seconds, the volume loss is predominantly constituted by the evaporation of ethanol (green dash-dotted line), while in the following, the evaporation of water (blue dashed line) determines the volume loss rate.}
\label{fgr:numpics}
\end{figure*}

The model solves the coupled processes of multi-component evaporation, Stokes flow in the droplet driven by solutal and thermal Marangoni flow and convection-diffusion equations for the spatio-temporal liquid composition and the temperature. The composition-dependence of the liquid properties, i.e. mass density, viscosity, surface tension, diffusivity and thermodynamic activities, are taken into account. However, since these relations are not available for the ternary mixture, experimental data of binary water-ethanol mixtures \cite{Vazquez1995a,Gonzalez2007a,Parez2013a} have been fitted to model the composition-dependence. The thermodynamic activities of water and ethanol were calculated by AIOMFAC \cite{Zuend2008a,Zuend2011a}. Plots of the fitted relations and description of all used parameters can be found in supplementary material.

The evaporation rate for component $\nu$ ($\nu=\text{w,e}$ for water and ethanol, respectively) are calculated by solving the vapor diffusion equation $\partial_t c_{\nu}=D^{\text{vap}}_{\nu,\text{air}}\nabla^2c_{\nu}$ in the gas phase, where $D^{\text{vap}}_{\nu,\text{air}}$ is the vapor diffusivity of $\nu$ in air and $c_{\nu}$ is the vapor concentration, i.e. the partial density. With the ideal gas law, one can express the ambient water vapor concentration by
\begin{equation}
c_{\text{w},\infty}=H\frac{M_{\text{w}}p_{\text{w,sat}}(T_{\infty})}{R T_\infty}\, ,
\label{eqn:cinfty}
\end{equation}
where $H$ is the relative humidity, $R$ is the ideal gas constant, $T_{\infty}$ is the room temperature and $p_{\text{w,sat}}$ is the saturation pressure, which temperature-dependence given by the Antoine equation. There is no ambient ethanol vapor present, i.e. $c_{\text{e},\infty}=0$.  At the liquid-gas interface, the vapor-liquid equilibrium according to Raoult's law has to hold,
\begin{equation}
c_{\nu,\text{VLE}}=\gamma_\nu x_\nu \frac{M_{\nu}p_{\nu\text{,sat}}(T)}{R T} \, .
\label{eqn:cvle}
\end{equation}
By virtue of equation \ref{eqn:cvle}, the evaporation rates are coupled with the local drop composition at the interface via the liquid mole fraction $x_\nu$ and the activity coefficient $\gamma_\nu$ and furthermore with the local temperature $T$. From the diffusive vapor fluxes $J^{\text{gas}}_{\nu}=-D^{\text{vap}}_{\nu,\text{air}}\partial_n c_{\nu}$ at the interface, the mass transfer rates $j_{\text{w}}$ and $j_{\text{e}}$ can be determined from the coupled mass transfer jump conditions
\begin{equation}
j_{\nu} - J^{\text{gas}}_{\nu} - \frac{c_{\nu}}{\rho_{\text{air}}}\left( j_{\text{e}} + j_{\text{w}} \right)=0 \, .
\label{eqn:mtjcgas}
\end{equation}
Here, the mass density $\rho_{\text{air}}$ of the gas phase is assumed to be constant and given by the value of humid air.

The drop is assumed to be in a spherical-cap shape, where the base contact radius $r_{\text{c}}$ was adjusted according to the experimental data (Data-2), i.e. by fitting $r_{\text{c}}(V)$. This allows to calculate the normal interface velocity from the evaporative volume loss via the kinematic boundary condition. Together with the tangential Marangoni shear stress boundary condition, the Stokes flow can be solved in the drop. The obtained velocity $\vec{u}$ is entering the convection-diffusion equations for the composition, expressed in terms of mass fractions $y_{\nu}$, i.e. 
\begin{equation}
\rho\left(\partial_t y_{\nu} + \vec{u}\cdot \nabla y_{\nu}\right) = \nabla \cdot \left(\rho D\nabla y_{\nu}\right) -J_\nu \delta_{\Gamma} \,,
\label{eqn:massfraccde}
\end{equation}
with the mixture diffusivity $D$ and the mass transfer source/sink term imposed at the liquid-gas interface $\Gamma$. The source/sink term is given by
\begin{equation}
J_\nu=j_{\nu}-y_{\nu}\left( j_{\text{e}} + j_{\text{w}} \right) \, .
\label{eqn:mtjcliq}
\end{equation}
The mass fraction $y_{\text{ta}}$ of trans-anethole oil is obtained by $y_{\text{ta}}=1-y_{\text{w}}-y_{\text{e}}$.

Finally, the following temperature equation has to be solved:
\begin{equation}
\rho c_p \left(\partial_t T + \vec{u}\cdot \nabla T\right) = \nabla \cdot \left(\lambda\nabla T\right) -\left(\Lambda_{\text{w}} j_{\text{w}}+\Lambda_{\text{e}} j_{\text{e}}\right) \delta_{\Gamma} \,.
\label{eqn:tempcde}
\end{equation}
Here, $\Lambda_{\nu}$ is the latent heat of evaporation, $\rho$ is the mass density, $c_p$ the specific heat capacity and $\lambda$ is the thermal conductivity. These quantities are different in the different domains, i.e. gas phase, drop and substrate. In order to accurately reproduce the thermal conduction of the experimental setup, a fused quartz substrate with a finite thickness of \SI{1.17}{\milli\meter} and air below is considered. 

The simulation results confirm the interpretation discussed in section~\ref{sec:nucleEXP} on the phenomenon that the evaporation-triggered nucleation starts to occur at the top of an evaporating ouzo drop.
The snapshots of the simulation (Figs.~\ref{fgr:numpics}AB) shows that the ouzo effect indeed sets in close to the apex at about $t=\SI{20}{\second}$ (white regions in Fig.~\ref{fgr:numpics}A). 
At $t=\SI{26.3}{\second}$, phase separation is occurring in the entire drop.
Figure~\ref{fgr:numpics}C presents the temporal evolution of the drop volume and the partial volume of each component.
The two distinct slopes in the drop volume curve in the early stage and at later stages correspond to the curve slopes of ethanol and water, respectively.
Furthermore, the FEM simulation provides more information about the evaporating ouzo drop.
Due to the enhanced evaporation at the top of the drop, both the temperature and the ethanol concentration have their minimum close to the apex. Since both aspects increase the local surface tension, a strong Marangoni flow is induced from the contact line along the interface towards the top. As a result of the solutal Marangoni instability, however, the flow in the drop is, in general, not regular, but exhibits chaotic behavior until almost all ethanol has evaporated. Thus, one also has to expect axial symmetry breaking, which is discussed in more detail in a forthcoming three-dimensional investigation of the evaporating ouzo drop \cite{Diddens2017c}.

\section{Generalized Diffusion Model for the Evaporation Process of Ouzo Drops }
\label{sec:model}

\begin{figure*}[ht]
 \centering
 \includegraphics[width=.9\linewidth]{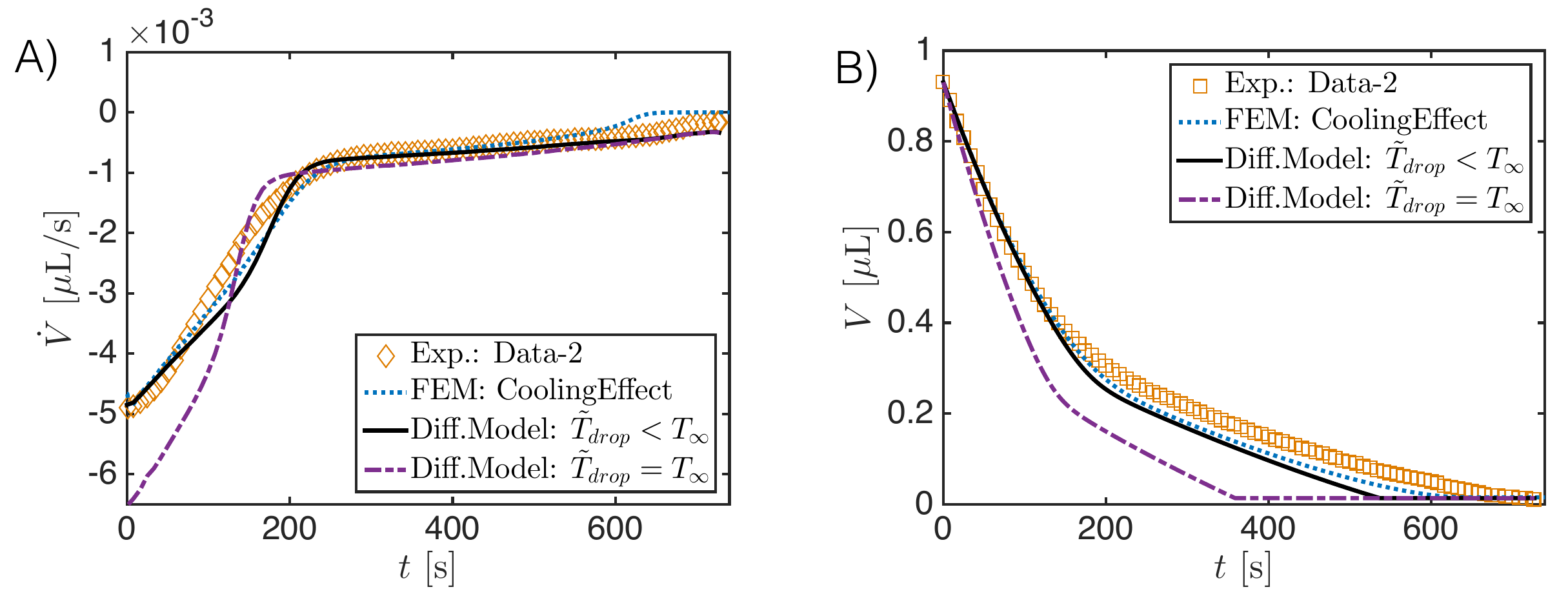}
 \caption{Evolutions of volume loss rate (panel A) and drop size (panel B) calculated from the approximate diffusion model ($\tilde{T}_{\text{drop}}<T_{\infty}$), displayed in black solid lines, and the model without cooling-effect consideration ($\tilde{T}_{\text{drop}}=T_{\infty}$) , presented in purple dash-dotted lines.
The blue dotted lines are the results from the FEM simulation (cooling effect included).
The model results are in a good accordance with experimental data and FEM simulation results.
When the approximate diffusion model excludes the evaporative cooling effect, the satisfaction to the experimental data and simulation results lose, and the calculated volume loss rates are higher, especially in the early stage where ethanol evaporation dominates.}
 \label{fgr:model}
\end{figure*}
 
As stated above, an analytical diffusion model for quasi-steady natural evaporation of one-component drops was proposed by Popov \cite{popov2005}.  
To generalize Popov's diffusion model for the evaporation process of ouzo drops with more than one component, we take account of Raoult's law, which is necessary for building up the vapor-liquid equilibrium at the drop interface \cite{tanouzo2016,Diddens2017a}.
We assume the liquid solution in the drop to be well-mixed as a result of the existence of the strong Marangoni flow  \cite{tanouzo2016,Diddens2017c}.
The mixed-convection flow gives rise to both a uniform concentration field and a uniform thermal field inside the drop, which is discussed in detail in section~\ref{sec:marangoni}.
In the generalized model, the temperature dependence of the vapor concentration of each component is also considered, because evaporative cooling effects at the drop interface is obvious for an evaporating drop at a high contact angle \cite{dash2013}. Detailed discussion is presented in section~\ref{sec:coolingeffect}.

\subsection{Generalized Diffusion Model}
In Popov's model, the evaporation flux $J(r)$ on the surface of a pure fluid drop is given by \cite{popov2005}
\begin{eqnarray}
& J(r)=\frac{D^{\text{vap}}_{\text{air}}(c_{\text{sat}}-c_{\infty})}{r_c}\Bigg[\frac{1}{2}\sin{\theta}+\sqrt{2}(\cosh{\alpha}+\cos{\theta})^{3/2}\nonumber \\
& \times  \int_{0}^{\infty}\frac{\cosh{\theta\tau}}{\cosh{\pi\tau}}\tanh{[(\pi-\theta)\tau]}P_{-1/2+i\tau}(\cosh{\alpha})\tau \diff \tau \Bigg]\, , 
\label{eqn:popJ}
\end{eqnarray}
where $r=\frac{r_{c}\sinh{\alpha}}{\cosh{\alpha}+\cos{\theta}}$ is the radial coordinate at the surface of the drop,  $r_c$ is the contact radius of the drop, $D^{\text{vap}}_{\text{air}}$ is the coefficient of vapor diffusion, $c_{\text{s}}$ is the saturated vapor concentration on the drop surface, $c_{\infty}$ is the concentration of vapor at infinity, and $\theta$ is contact angle (cf. sketch in Fig.~2E).

To generalize the model for ouzo drops, the evaporation flux along the vapor-liquid interface of each component $\nu$ can be expressed as
\begin{eqnarray}
& J_{\nu}(r)=\frac{D^{\text{vap}}_{\nu,\text{air}}(c_{\nu,\text{VLE}}-c_{\nu,\infty})}{r_c}\Bigg[\frac{1}{2}\sin{\theta}+\sqrt{2}(\cosh{\alpha}+\cos{\theta})^{3/2}\nonumber \\
& \times  \int_{0}^{\infty}\frac{\cosh{\theta\tau}}{\cosh{\pi\tau}}\tanh{[(\pi-\theta)\tau]}P_{-1/2+i\tau}(\cosh{\alpha})\tau \diff \tau\Bigg]\, \nonumber, 
\label{eqn:popJ}
\end{eqnarray}
where the concentration of each component at the vapor-liquid interface is determined by Raoult's law, i.e. equation~(2), $c_{\nu,\text{VLE}}=\gamma_{\nu}x_{\nu}c_{\nu,\text{sat}}$.

The evaporation rate of the mass of each component $\dot{m}_{\nu}$ is expressed as an integral of the evaporation flux over the drop surface, i.e. $\dot{m}_{\nu}=-\int_{0}^{r_c}J_{\nu}(r) \sqrt{1+(\partial_{r}h(r))^2}2\pi r \diff r$.
By virtue of the assumption of uniform concentration and homogeneous temperature along the interface, both vapor concentration $c_{\nu,\text{VLE}}$ at the drop surface and vapor diffusivity $D^{\text{vap}}_{\nu,\text{air}}$ are independent of position coordinate $r$. 
The concentration $c_{\nu,\text{VLE}}$ depends on the averaged surface temperature $\tilde{T}_{\text{drop}}$ of the drop.
Both the diffusivity $D^{\text{vap}}_{\nu,\text{air}}$ and the ambient vapor concentration $c_{\nu,\infty}$ are determined by room temperature $T_{\infty}$.
Therefore, the evaporation rate of drop mass $\dot{m}$ can be expressed as follows:
 \begin{eqnarray}
& \dot{m}_{\nu}=-\pi r_{c}D^{\text{vap}}_{\nu,\text{air}}\Bigg(c_{\nu,\text{VLE}}(\tilde{T}_{\text{drop}})-c_{\nu,\infty}(T_{\infty})\Bigg)f(\theta)\, , \\
& c_{\nu,\text{VLE}}(\tilde{T}_{\text{drop}})=\gamma_\nu(\tilde{T}_{\text{drop}}) x_\nu \frac{M_{\nu}P_{\nu\text{,sat}}(\tilde{T}_{\text{drop}})}{R \tilde{T}_{\text{drop}}}\, , \nonumber \\
& f(\theta)=\frac{\sin{\theta}}{1+\cos{\theta}}+4\int_{0}^{\infty}\frac{1+\cosh{2\theta\tau}}{\sinh{2\pi\tau}}\tanh{[(\pi-\theta)\tau]}\diff \tau\, . \nonumber
 \label{eqn:popsim}
 \end{eqnarray}
The evaporation rate of the drop mass $\dot{m}$ is given by $\dot{m}=\sum\dot{m}_{\nu}=\sum\rho_{\nu}\dot{V}_{\nu}$, where $\dot{V}_{\nu}$ is the evaporation rate of each component volume. 
When the small volume-change caused by the mixture of miscible liquids is ignored, the evaporation rate of the drop volume $\dot{V}$ can be expressed as,
\begin{equation}
\dot{V}=\sum\frac{1}{\rho_{\nu}}\dot{m}_{\nu}\, .
\label{eqn:dV}
\end{equation}

In the application of the generalized diffusion model for the evaporation process of an ouzo drop, we assume that the trans-anethole ($\nu=\text{ta}$) is non-volatile and has no influence on the evaporation of water ($\nu=\text{w}$) and ethanol ($\nu=\text{e}$). 
For water ($\nu = \text{w}$), the ambient vapor concentration $c_{\text{w},\infty}$ is given by $H c_{\text{w,sat}}(T_\infty)$, i.e. equation~\ref{eqn:cinfty}.
For ethanol ($\nu = \text{e}$), we assume that the ambient vapor concentration $c_{\text{e},\infty}$ is zero, thus its evaporation rate is independent of the $H$ factor.
The saturation pressure $P_{\text{w,sat}}$ and $P_{\text{e,sat}}$ in equations~\ref{eqn:cinfty} and \ref{eqn:cvle} at different temperatures are calculated from the Antoine equation.
The activity coefficients $\gamma_{\nu}$ are calculated from the \textit{Dortmund Data Bank} with the lastest available parameters \cite{wittig2003vapor}.
Vapor diffusivity $D^{\text{vap}}_{\text{w, air}}$ at different temperatures is obtained from reference  \cite{lide2004crc} by cubic spline interpolation.
The vapor diffusivity $D^{\text{vap}}_{\text{e, air}}$ is calculated based on the equation in reference \cite{Lapuerta2014}.
We use $\rho_{w}=$ \SI{997.773}{\kilogram\per\cubic\metre}, $\rho_{e}=$ \SI{786.907}{\kilogram\per\cubic\metre}, $M_{w}=$ \SI{0.018}{\kilogram\per\mole} and $M_{e}=$ \SI{0.046}{\kilogram\per\mole}.
The contact radius of the drop $r_c$, the contact angle $\theta$, the ambient temperature $T_{\infty}$ and the relative humidity $H$ at each moment were measured in the experiments. 
The decreased drop temperature $\tilde{T}_{\text{drop}}$ is substituted by the interface-averaged temperature from the FEM simulation in section~\ref{sec:num}.
The initial mole fractions of water $x_{\text{w}}$ and ethanol $x_{\text{e}}$ are calculated based on the initial water and ethanol composition of our ouzo liquid.
Then $x_{\text{w}}$ and $x_{\text{e}}$ are recalculated in each step base on volume loss rate $\dot{V}_{i}$ and their values in the previous step.
The calculation was performed in MATLAB with homemade codes.

Figure~\ref{fgr:model}A shows the evaporation rate of the drop volume predicted by the generalized diffusion model (black solid line) has a good agreement both with the measured experiment data (diamond points) and with FEM simulation results (blue dotted line). 
However, when the evaporative cooling effect is ignored in the model (purple dash-dotted line), i.e. $\tilde{T}_{\text{drop}}=T_{\infty}$, a clear  deviation of the evaporation rates appears, especially at early stage where the ethanol evaporation dominates.
This is a result of the high volatility of ethanol which enhances the evaporative cooling effect.
At later stages, the deviation disappears caused by two factors.
The first factor is the low volatility of water, which reduces the evaporative cooling effect.
 The second factor is the decreased contact angle of the evaporating drop, as shown in figure~\ref{fgr:evp}B.
A smaller contact angle gives a smaller temperature reduction at the drop top \cite{dash2013} and also weakens the cooling effect.
An integral of the evaporation rate from initial time $t_0$ gives the temporal evolution of the drop volume, that is $V(t)=\int_{t_0}^{t}\dot{V} \diff t$.
As displayed in figure~\ref{fgr:model}B, the diffusion model provides a comparable evolution curve (black solid line) to the one (square points) from experimental data and the one (blue dotted line) from FEM simulation result.
There is slight deviation in the later stage, although the slopes of evolutionary curves are almost the same.
There are two potential reasons for this deviation: (i) The isothermal assumption doesn't satisfy strictly because of the reduced intensity of the Marangoni flow. (ii) The existence of the nucleated oil microdroplet in the drop plays a role.
For the case without considering the cooling effect, the temporal evolution curve (purple dash-dotted line) obviously deviates from the other curves.
The curve still develops with two different slopes, but the entire evaporation is predicted too fast.
It indicates that the evaporative cooling effect is a very important aspect in the evaporation process of the droplet with a high contact angle.

\subsection{Strong Marangoni flow}
\label{sec:marangoni}

For an evaporating multi-component drop, the surface tension gradient along the drop surface may be very small, whereas the Marangoni flow can be very strong \cite{tanouzo2016,Diddens2017a}.
For a millimetre water-ethanol drop, the averaged flow velocity inside is of order of \SI[per-mode=symbol]{}{\milli\metre\per\second} \cite{Diddens2017a,Diddens2017b}.
Its Péclet numbers $Pe$ for mass transfer and heat transfer are $10^5$ and $10^3$, respectively, given its mass diffusion coefficient ($\sim 10^{-9}$\SI{}{\square\metre\per\second}) and thermal diffusivity ($\sim10^{-7}$\SI{}{\square\metre\per\second}).
Although the inconsistent evaporation flux along the drop surface and the evaporative cooling effect causing the concentration gradient and thermal gradient, respectively, the strong Marangoni flow can dramatically uniformize both the concentration and thermal distribution in the drop.
Here we assume that the drop has both a uniform concentration field and a uniform thermal field inside. 
Then both the vapor concentration $c_{\nu,\text{VLE}}$  at the drop surface and vapor diffusivity $D^{\text{vap}}_{\nu,\text{air}}$ are independent of position coordinate $r$. 
The entire drop is characterized by an identical temperature value $\tilde{T}_{\text{drop}}$.
The concentration $c_{\nu,\text{VLE}}$ only depends on the drop (surface) temperature $\tilde{T}_{\text{drop}}$.

\subsection{Evaporative cooling effect at the drop surface}
\label{sec:coolingeffect}
Evaporative cooling effect is a very important aspect for the evaporation process of sessile drops\cite{girard2008,dash2013,pan2013,dash2014b,gleason2014}.
During the evaporation, the phase change at the liquid-air interface consumes energy and results in temperature reduction.
Meanwhile heat is replenished from the substrate by heat conduction.
These two main factors lead to a non-isothermal field in the drop \cite{dash2014b}.
There is a temperature distribution along the liquid-air interface.
The temperature difference within an evaporating drop on hydrophilic surfaces is minimal \cite{girard2008}, whereas there is a relative large temperature reduction across the drop on hydrophobic surfaces \cite{dash2013,pan2013,gleason2014}.
All the literature mentioned above are for a single-component drop.

In our case, the ouzo drop has a large contact angle.
It is vital to take account of the temperature reduction caused by evaporative cooling effects.
But something different happens here.
As discussed in section~\ref{sec:marangoni}, the appearance of the strong Marangoni flow uniformizes the temperature field.
It is possible to have a thermal boundary layer along the substrate surface.
The estimation of its thickness is $\sim$ \SI{100}{\micro\metre}, given by $Pe=1$.
And as a result there is no distinct temperature difference in most parts of the drop, as displayed in Figures~\ref{fgr:numpics}AB.
Therefore it is reasonable to assume an isothermal drop with a reduced temperature value.

\section*{Conclusions}
The evaporation of an ouzo drop on a superamphiphobic surface is characterized by three features: (i) Nucleation of oil microdroplets triggered at the top of the drop. (ii) Two distinct regimes
in the evaporation rates can be distinguished, with the formed oil wrapping around the ouzo drop. (iii) In the final stage of
evaporation a continuous oil phase cloaks the drop.
Quantitative results for the temporal evolution of the drop volume, contact angle and the size of contact area are presented.
A numerical simulation with a new FEM method \cite{Diddens2017b} verifies the evaporation-triggered nucleation at the drop top and provides additional insight into the entire physical process.
Although the inconsistent evaporation flux along the drop surface and the evaporative cooling effect cause the concentration gradient and thermal gradient, respectively, the Marangoni flow dramatically equalizes both the concentration and thermal gradients in the drop.
Taking advantage of the uniformity inside the drop, we propose a generalized diffusion model for the evaporation process of an ouzo drop based on Popov's theory.
We generated a model by integrating Raoult's law and the evaporative cooling effect and then simplified the model with the uniformity assumption. 
The proposed model provides a simplified way to analyze the evaporation process of multi-component drops. 
With the experimental data and FEM simulation results the predicted instantaneous volume of the ouzo drop shows good agreement.
It is appropriate to build up a proper model for the ouzo drop temperature distribution to complete the model.
This work highlights the influence of substrate on the evaporation process of ouzo drops.
A better understanding of the dynamics of an evaporating ouzo drop may provide valuable information for the investigation of the evaporation process of multi-component mixture drops.
 

\begin{addendum}
 \item We thank Alvaro Marin for the suggestion on the selection of the substrate, Pengyu Lv for the technological help on the confocal microscope system and Hans Kuerten for discussion on the numerics. 
We also thank Gabriele Schäfer, Florian Geyer and Sanghyuk Wooh from \textit{Max Planck Institute for Polymer Research} for the fabrication and shipment of the soot-templated superamphiphobic substrates.
H.T. thanks for the financial support from the China Scholarship Council (CSC, file No. 201406890017). C.D. gratefully acknowledge financial support by Océ -- A Canon Company and by the Dutch Technology Foundation STW, which is part of the Netherlands Organisation for Scientific Research (NWO), and which is partly funded by the Ministry of Economic Affairs. We also acknowledge the Dutch Organization for Research (NWO) and the Netherlands Center for Multiscale Catalytic Energy Conversion (MCEC) and the \textit{Max Planck Center Twente for Complex Fluid Dynamics} for financial support.
 \item[Competing Interests] The authors declare that they have no
competing financial interests.
 \item[Correspondence Email:] h.tan@utwente.nl or d.lohse@utwente.nl or xuehua.zhang@rmit.edu.au.
\end{addendum}


\clearpage
\newpage
\renewcommand{\thefigure}{S.\arabic{figure}}    
\renewcommand{\theequation}{S.\arabic{equation}}
\renewcommand{\thesection}{S.\arabic{section}}
\renewcommand{\thetable}{S.\arabic{table}}    

\section*{Supporting Information}
\section{Phase diagram of the trans-anethole-ethanol-water system}
Figure~\ref{fig:tg} is the ternary diagram of the trans-anethole-ethanol-water system. The blue solid line is the measured phase-separation curve. The gray dashed lines indicate the composition paths of the titration experiments. The titration was conducted at a temperature of around \SI{22}{\celsius}.

\section{Parameters used in the FEM model}

\subsection{List of symbols}
A description of all symbols used in the FEM model can be found in Table~\ref{tab:si:listofsymbols}. For quantities that are constant during the simulation, also the corresponding values are given.
\begin{table*}
\tiny
\centering
\begin{tabular}{|c|l|c|}
\hline 
\textbf{Symbol} & \textbf{Description} & \textbf{value and/or unit} \\
\hline
$c_\nu$ & vapor concentration of species $\nu$ & $\si{\kilogram\per\meter^3}$\\
$c_{\nu,\infty}$ & ambient vapor concentration of species $\nu$ & $\si{\kilogram\per\meter^3}$\\
$c_{\nu,\text{VLE}}$ & vapor-liquid equilibrium concentration & $\si{\kilogram\per\meter^3}$\\
$c_{p}$ & specific heat capacity & $\si{\joule\per(\kilogram\kelvin)}$\\
$D$ & mutual diffusivity in the liquid & $\si{\meter^2\per\second}$\\
$D^{\text{vap}}_{\nu,\text{air}}$ & vapor diffusion coefficient & $D^{\text{vap}}_{\text{w},\text{air}}=\SI{0.260}{\centi\meter^2\per\second}$ \cite{Lee1954a}\\
 &  & $D^{\text{vap}}_{\text{e},\text{air}}=\SI{0.135}{\centi\meter^2\per\second}$ \cite{Lee1954a}\\
$H$ & relative humidity of water & \SI{42}{\percent}\\
$j_\nu$ & mass transfer rate of species $\nu$ & $\si{\kilogram\per(\meter^2\second)}$\\
$J_\nu$ & diffusive liquid flux at the interface & $\si{\kilogram\per(\meter^2\second)}$\\
$J^{\text{gas}}_\nu$ & diffusive vapor flux at the interface & $\si{\kilogram\per(\meter^2\second)}$\\
$M_\nu$ & molar mass of species $\nu$ & $M_{\text{w}}=\SI{18.015}{\gram\per\mol}$ \cite{DortmunderDatenbank}\\
 &  & $M_{\text{e}}=\SI{46.069}{\gram\per\mol}$ \cite{DortmunderDatenbank}\\
 &  & $M_{\text{ta}}=\SI{148.21}{\gram\per\mol}$ \cite{PubChemAnethole}\\
$p_{\nu,\text{sat}}$ & saturation pressure of component $\nu$ & $\si{\pascal}$\\
$r_\text{c}$ & base radius & \si{\meter}\\ 
$R$ & universal gas constant & $\SI{8.3144598}{\joule\per(\mol\kelvin)}$\\
$t$ & time & $\si{\second}$\\
$T$ & temperature & \si{\celsius}\\
$T_{\infty}$ & ambient temperature & \SI{23}{\celsius}\\
$\vec{u}$ & mass-averaged liquid velocity & $\si{\meter\per\second}$\\
$V$ & droplet volume & \si{\meter^3}\\ 
$x_{\nu}$ & mole fraction of component $\nu$ in the liquid & \\
$y_{\nu}$ & mass fraction of component $\nu$ in the liquid & \\
$\gamma_{\nu}$ & activity coefficient of component $\nu$ & \\
$\delta_{\Gamma}$ & delta function at the liquid-air interface & \si{1\per\meter^2}\\
$\theta$ & contact angle & \si{\degree}\\ 
$\lambda$ & thermal conductivity & $\si{\watt\per(\meter\kelvin)}$\\
$\Lambda_{\nu}$ & latent heat of evaporation & $\Lambda_{\text{w}}=\SI{2438}{\kilo\joule/\kilo\gram}$ \cite{DortmunderDatenbank}\\
 &  & $\Lambda_{\text{e}}=\SI{918}{\kilo\joule/\kilo\gram}$ \cite{DortmunderDatenbank}\\
$\nu$ & component index & $\nu=\text{w},\text{e},\text{ta}$\\  
$\rho$ & mass density & $\si{\kilogram\per\meter^3}$\\
$\rho^{\text{gas}}$ & mass density of air & $\SI{1.183}{\kilogram\per\meter^3}$ \cite{bookDean1992a}\\
\hline
\end{tabular}
\label{tab:si:listofsymbols}
\caption{A table containing all quantities entering the FEM model. If the quantity is constant during the simulation, also the corresponding value is given.}
\end{table*}

\subsection{Relations for non-constant quantities}
\subsubsection{Saturation pressure $p_{\nu,\text{sat}}$}
The temperature-dependence of the saturation pressure is calculated by the Antoine equation, i.e. by
\begin{equation}
\operatorname{log}_{10}\left(p_{\nu,\text{sat}} [\text{in \si{\mmHg}}]\right)=A_\nu-\frac{B_\nu}{C_\nu+T[\text{in \si{\celsius}}]}\,,
\end{equation}
where the constants $A_\nu$, $B_\nu$ and $C_\nu$ read \cite{DortmunderDatenbank}
\begin{equation}
\nonumber
\begin{tabular}{r|ccc|}
  & $A_\nu$ & $B_\nu$ & $C_\nu$ \\
\hline 
water & 8.07131 & 1730.63 & 233.426 \\
ethanol & 8.20417 & 1642.89 & 230.300 \\
\hline 
\end{tabular}
\end{equation}

\subsubsection{Composition-dependent properties}
In the droplet, the physical properties depend on the mixture composition. Due to the low initial concentration of trans-anethole, the composition-dependence of all quantities in the ouzo droplet was approximated based on a binary water-ethanol mixture. To that end, experimental data for the mass density $\rho$ \cite{Gonzalez2007a}, the dynamic viscosity $\mu$ \cite{Gonzalez2007a}, the surface tension $\sigma$ \cite{Vazquez1995a}, the diffusivity $D$ \cite{Parez2013a}, the specific heat capacity $c_p$ \cite{Grolier1981a} and the thermal conductivity $\lambda$ \cite{Yano1988a} was fitted. The activity coefficients $\gamma_\nu$ were determined by \textsl{AIOMFAC} \cite{Zuend2008a,Zuend2011a}. The extracted experimental data and the corresponding fits are depicted in \figurename~\ref{fig:params:propfitswe}.

\subsubsection{Thermal properties of air and substrate}
The thermal properties of the gas phase and the substrate used in the simulation read
\begin{equation}
\nonumber
\begin{tabular}{rl|ccc|}
 & &  $\rho$ [\si{\kilogram\per\meter^3}] & $c_p$ [\si{\joule\per(\kilogram\kelvin)}] & $\lambda$ [\si{\watt\per(\meter\kelvin)}]\\
\hline 
gas phase &(air) & 1.183 \cite{bookDean1992a} & 1005 \cite{Hilsenrath2013a} & 0.026 \cite{bookDean1992a} \\
substrate &(quartz glass) & 2648  \cite{bookDean1992a} & 739 \cite{bookDean1992a} & 1.36 \cite{Sergeev1982a} \\
\hline 
\end{tabular}
\end{equation}

\section{Temperature and relative humidity}
Figure~\ref{fig:RHT} presents the temperature $T_{\infty}$ and relative humidity $H$  in the laboratory during the experiments.
The sampling rate is one per second with a relative humidity accuracy of $\pm$ \SI{2}{\percent} over 10 to \SI{90}{\percent} @\SI{25}{\degreeCelsius} and a temperature accuracy  of $\pm$ \SI{0.3}{\kelvin} @\SI{25}{\degreeCelsius}).

\setcounter{figure}{0}
\begin{figure*}[ht]
\centering
  \includegraphics[width=.6\linewidth]{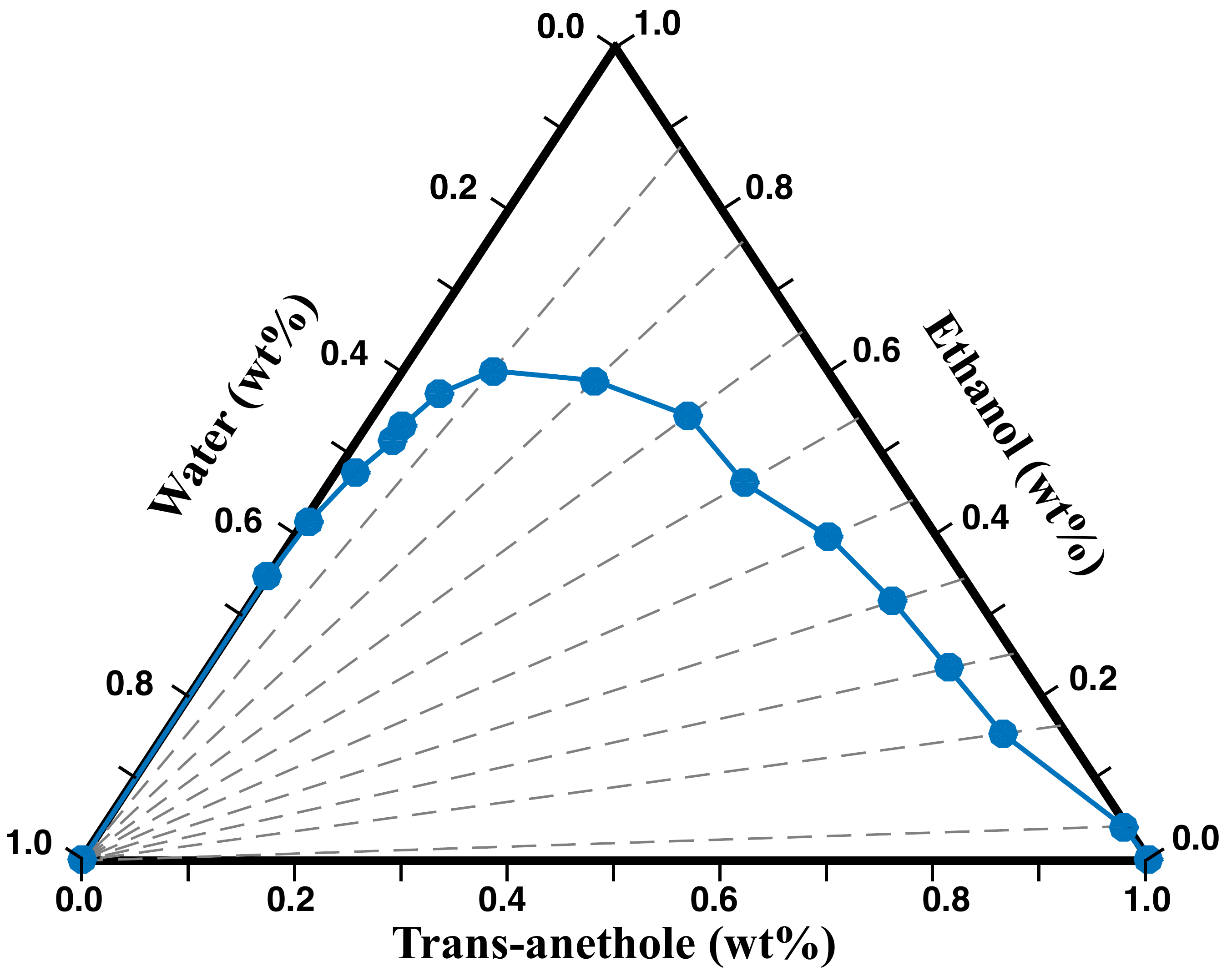}
\caption{Phase diagram for the trans-anethole-ethanol-water system. The blue dots present the measured miscibility limit.}
\label{fig:tg}       
\end{figure*}

\begin{figure*}[ht] 
\centering   
\includegraphics[width=1\textwidth]{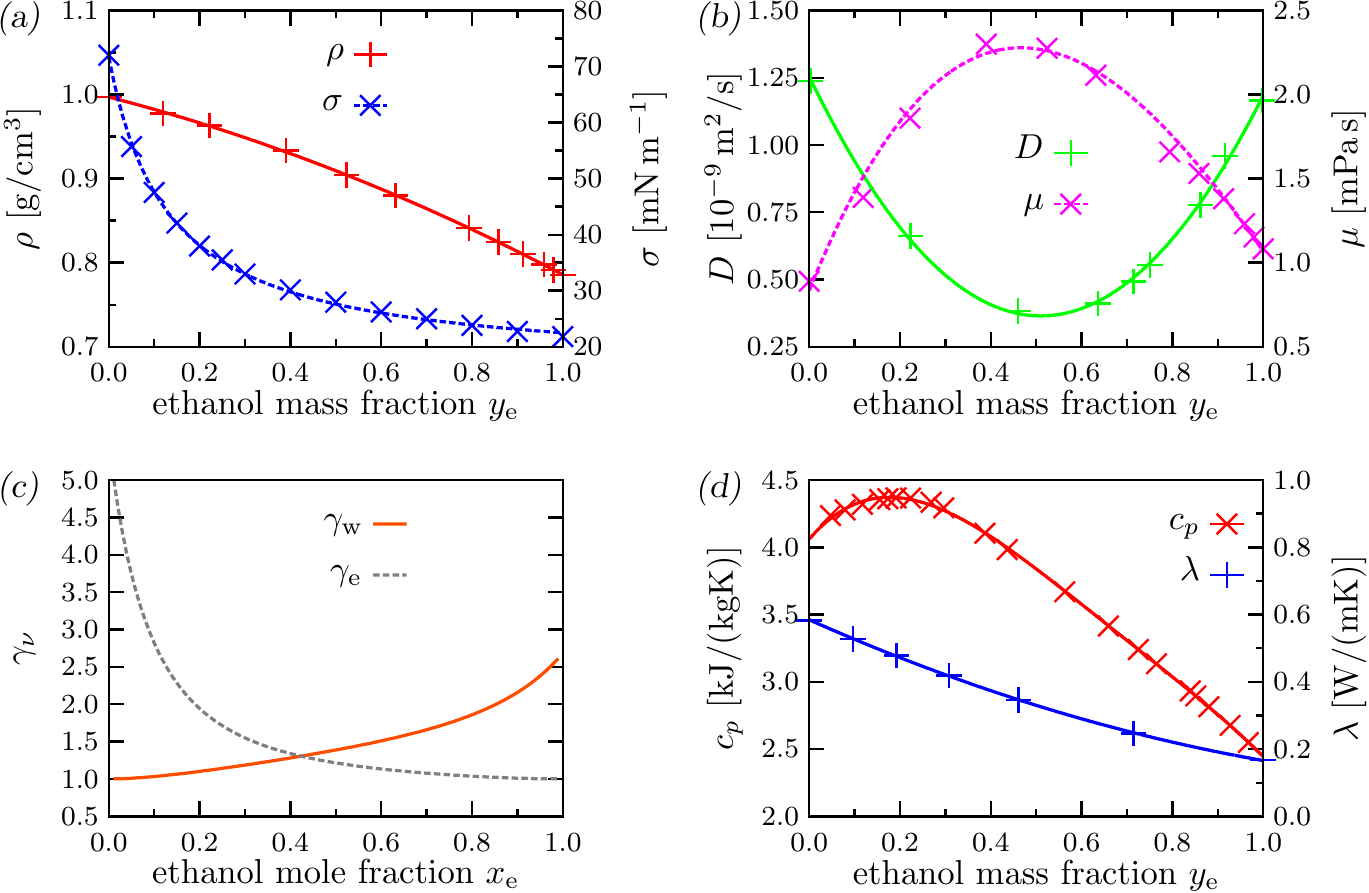} 
\caption{Composition-dependent properties based on a water-ethanol mixture. (a) Mass density $\rho$ \cite{Gonzalez2007a} and surface tension $\sigma$ \cite{Vazquez1995a}. Note that the surface tension is also a function of the temperature. However, since the temperature-dependence of the surface tension is only in the order of $\SI{-0.14}{\milli\newton\meter^{-1}\kelvin^{-1}}$, it is not plotted for the sake of visibility. (b) Dynamic viscosity $\mu$ \cite{Gonzalez2007a} and diffusivity $D$ \cite{Parez2013a}. (c) Activity coefficients $\gamma_\nu$ calculated by \textsl{AIOMFAC} \cite{Zuend2008a,Zuend2011a}. (d) Specific heat capacity $c_p$ \cite{Grolier1981a} and thermal conductivity $\lambda$ \cite{Yano1988a}. } \label{fig:params:propfitswe}  
\end{figure*}

\begin{figure*}[ht]
\centering
\begin{minipage}{0.5\textwidth} 
\includegraphics[width=1\textwidth]{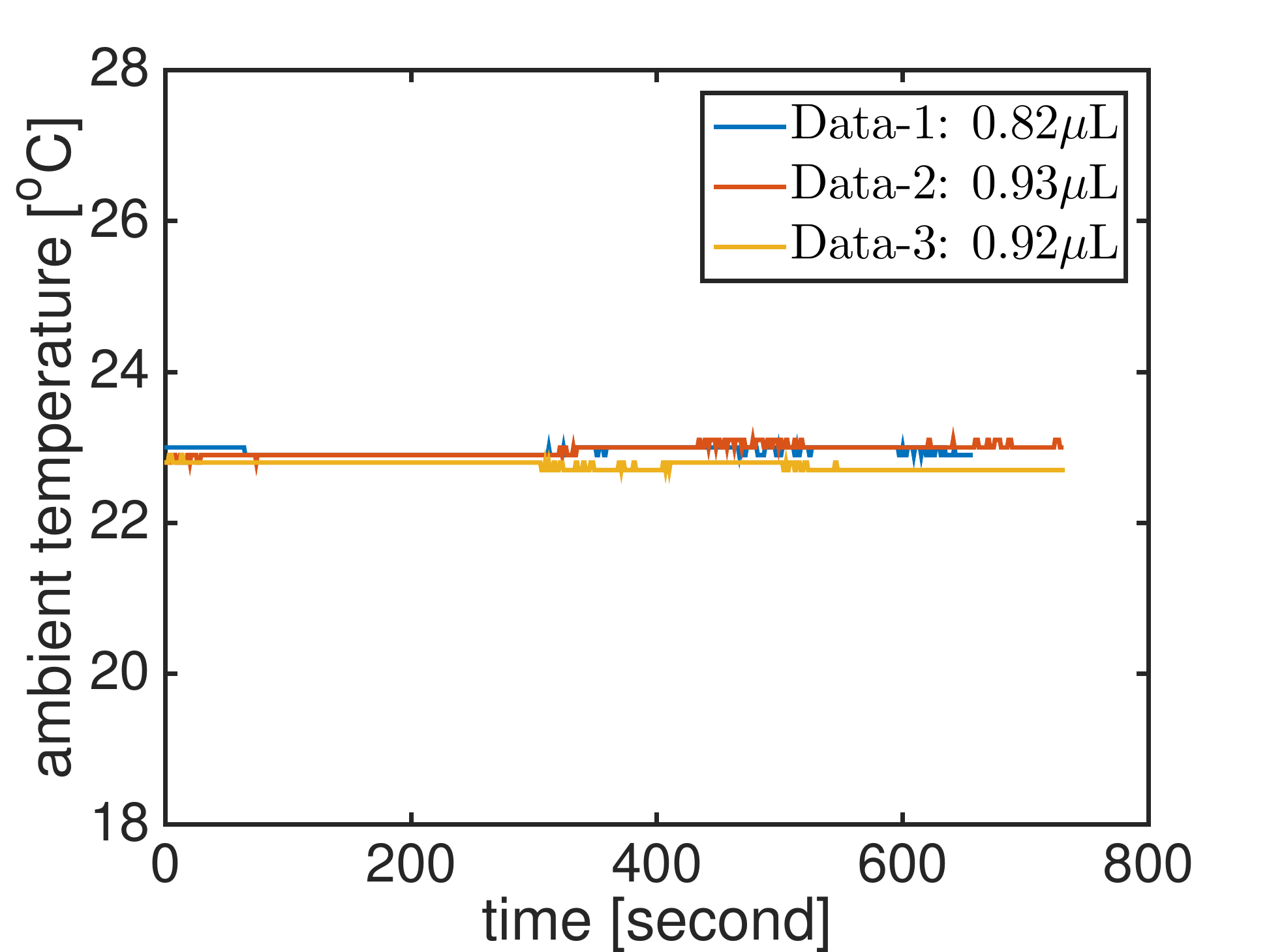}
\end{minipage}\hfill
\begin{minipage}{0.5\textwidth} 
\includegraphics[width=1\textwidth]{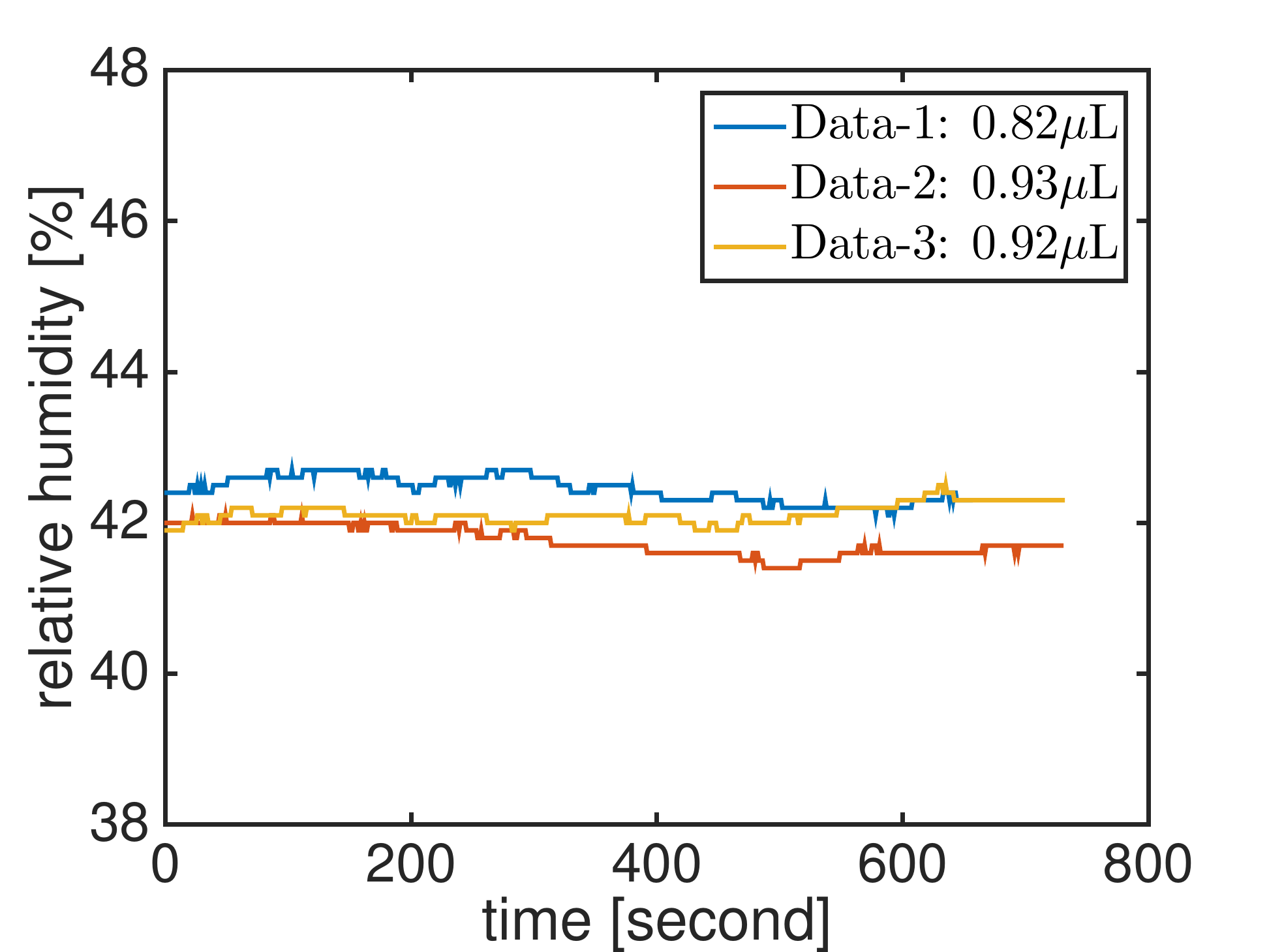}
\end{minipage}\hfill
\caption{Measured ambient temperature $T_{\infty}$ and relative humidity $H$ during the evaporation experiments.}
\label{fig:RHT}  
\end{figure*} 

\clearpage
\newpage
\section*{Supporting Information Reference}

\end{document}